\documentclass[journal]{IEEEtran}
\usepackage{cite}
\usepackage{bm}
\usepackage{multirow}
\usepackage[cmex10]{amsmath}
\usepackage{graphicx}
\usepackage{url}
\usepackage{color}
\usepackage{amsfonts}
\usepackage{amsmath}
\usepackage{extarrows}
\usepackage{graphicx}
\usepackage{amsfonts}
\usepackage{caption}
\usepackage{algorithm}
\usepackage{algpseudocode}
\usepackage{indentfirst}
\usepackage{caption}
\usepackage{esint} 
\usepackage{graphicx}
\usepackage{graphics}
\usepackage{epsfig}
\usepackage{soul}
\usepackage{caption}
\usepackage{gensymb}
 \DeclareMathOperator*{\argmax}{argmax}
\DeclareMathOperator*{\argmin}{argmin}

\makeatletter
\def\endthebibliography{%
  \def\@noitemerr{\@latex@warning{Empty `thebibliography' environment}}%
  \endlist
}
\makeatother

   \def\bD{{\mathbf{D}}} 
\def\bF{{\mathbf{F}}}  \def\bH{{\mathbf{H}}}  \def\bJ{{\mathbf{J}}}
    
    \def\bT{{\mathbf{T}}}

 \def\bb{{\mathbf{b}}}   
\def\bf{{\mathbf{f}}}    
 \def\bl{{\mathbf{l}}} \def\bm{{\mathbf{m}}} \def\bn{{\mathbf{n}}} 
    
  \def\bw{{\mathbf{w}}} \def\bx{{\mathbf{x}}} \def\by{{\mathbf{y}}}

\def\tcr{\textcolor{black}}

\def\argmin{\mathop{\mathrm{argmin}}}

\begin{document}
\title{Large Intelligent Surface for Positioning in Millimeter Wave MIMO Systems}
\author{Jiguang~He$^\dag$, Henk~Wymeersch$^\star$, Long Kong$^\ast$, Olli Silv\'en$^\ddag$, Markku~Juntti$^\dag$\\
        $^\dag$Centre for Wireless Communications, FI-90014, University of Oulu, Finland\\
        $^\star$Department of Electrical Engineering, Chalmers University of Technology, Gothenburg, Sweden\\
        $^\ast$Interdisciplinary Centre for Security Reliablility and Trust (SnT), University of Luxembourg, Luxembourg\\
        $^\ddag$Center for Machine Vision and Signal Analysis (CMVS), FI-90014, University of Oulu, Finland\\
\thanks{This work has been performed in the framework of the IIoT Connectivity for Mechanical Systems (ICONICAL), funded by the Academy of Finland. This work is also partially supported by the Academy of Finland 6Genesis Flagship (grant 318927) and Swedish Research Council (grant no. 2018-03701). }}
 \maketitle
\begin{abstract}
Millimeter-wave (mmWave) multiple-input multiple-output (MIMO) system for the fifth generation (5G) cellular communications can also enable single-anchor positioning and object tracking due to its large bandwidth and inherently high angular resolution. In this paper, we introduce the newly invented concept, large intelligent surface (LIS), to mmWave positioning systems, study the theoretical performance bounds (i.e., Cram\'er-Rao lower bounds) for positioning, and evaluate the impact of the number of LIS elements and the value of phase shifters on the position estimation accuracy compared to the conventional scheme with one direct link and one non-line-of-sight path. It is verified that better performance can be achieved with a LIS from the theoretical analyses and numerical study. 
\end{abstract}

\section{Introduction}
Conventionally, indoor and outdoor positioning is carried out by using received signal strength (RSS), time-difference-of-arrival (TDoA)~\cite{Yassin2017} or fingerprinting-based approaches~\cite{Savic2015,Wang2017}. Recently, efforts on positioning has been made by leveraging millimeter-wave (mmWave) multiple-input multiple-output (MIMO) systems and the geometric relationship between the base station (BS) and the mobile station (MS)~\cite{Shahmansoori2018, Abu-Shaban2018,Zhao2018, Koirala2018,Kakkavas2019}. It was shown that even a single BS can achieve promising positioning accuracy. Extensions to multiple-carrier and multi-user scenarios were studied in~\cite{Koirala2018}. 

Practical positioning algorithms for the single-anchor mmWave MIMO system can be classified into two major categories, i.e., direct positioning~\cite{Zhao2018} and two-stage positioning~\cite{Shahmansoori2018, Abu-Shaban2018, Koirala2018,Kakkavas2019}. The direct positioning aims at estimating the coordinates of the MS from the received signal directly, while the two-stage positioning first estimates the instantaneous channel realization: the channel gains, the angle of departure (AoD), the angle of arrival (AoA), and the time of arrival (ToA). In the second stage, the user location is calculated based on the channel estimates and the environmental geometry. 

Large intelligent surfaces (LIS), also known as reconfigurable intelligent surfaces (RIS), can effectively control the propagation wave, e.g., phase and even amplitude, without any need of baseband processing units~\cite{Huang2018, Ertugrul2019arXiv}. The LIS has been proposed to be used as a Tx/Rx antenna like in hybrid beamforming for positioning~\cite{Hu2018} and relay type reflector for communications~\cite{Huang2018}. 

In this paper, we study the positioning with the assistance of a LIS based reflector and multiple subcarriers at mmWave frequency bands. First, the performance bounds (i.e., Cram\'er-Rao lower bound) are evaluated based on the equivalent Fisher information matrix (FIM). The impact of the LIS, e.g., phase shifter value and the number of LIS elements, is studied on the estimation of channel parameters, positioning error bound (PEB), and orientation error bound (OEB). Numerical results show the superiority of the LIS aided mmWave MIMO positioning system over its conventional counterpart without incorporating a LIS. 
%

\section{System Model}
\begin{figure}[t]
	\centering
	\includegraphics[width=0.7\linewidth]{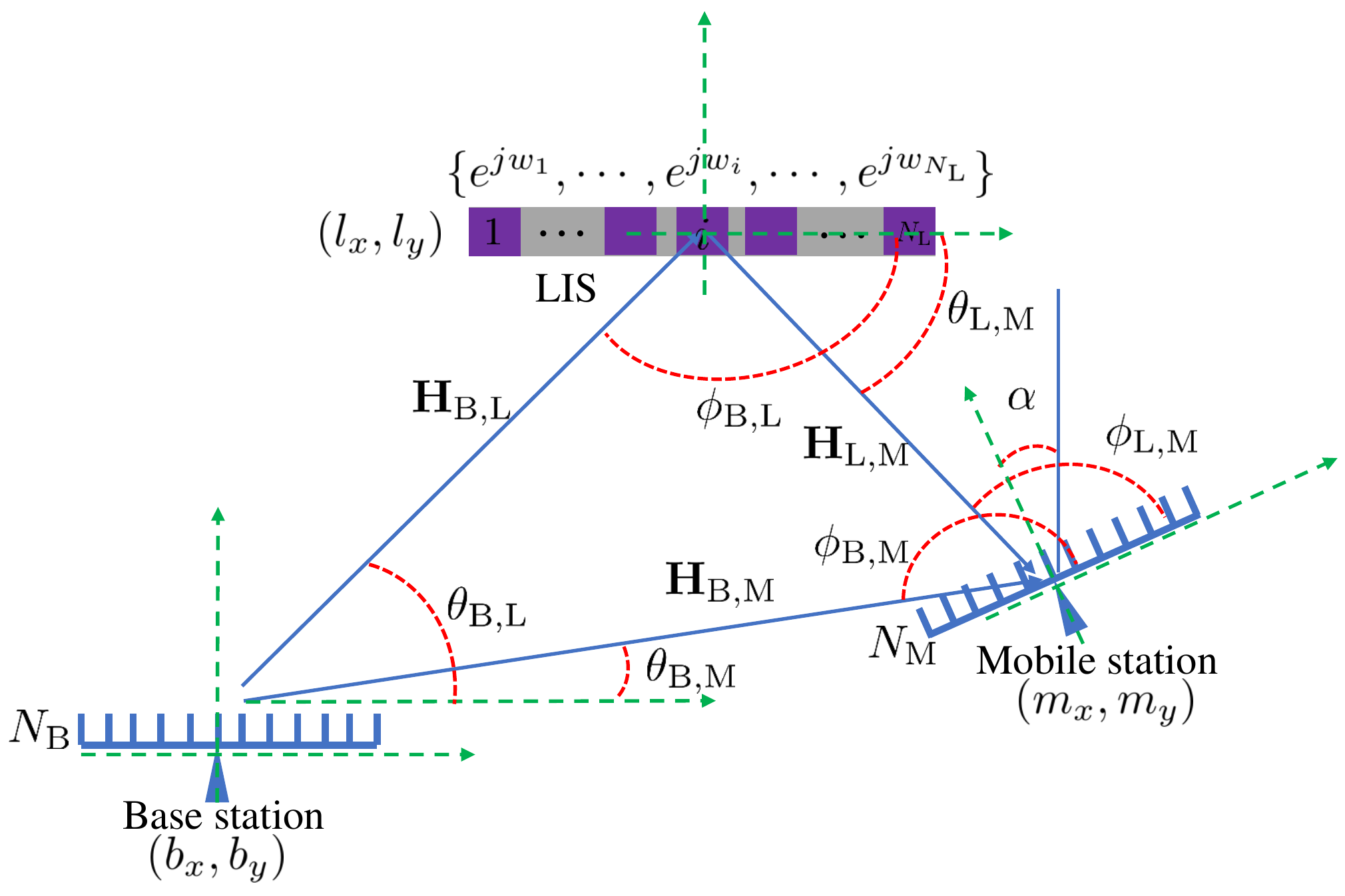}
	\caption{Positioning system with the aid of a large intelligent surface and multi-carrier mmWave OFDM signals. The coordinates and orientation of MS, $(m_x, m_y)$ and $\alpha$, are unknown and to be estimated.}
	\label{System_Model}
\end{figure}
The positioning system is presented in Fig.~\ref{System_Model}, which consists of one multiple-antenna BS, one multiple-antenna MS and one LIS. We consider a two-dimensional (2D) scenario with uniform linear arrays (ULAs) for both the antenna elements and LIS elements (i.e., analog phase shifters). The numbers of antenna elements at the BS and MS are $N_\text{B}$ and $N_\text{M}$, respectively, while the number of LIS elements is $N_\text{L}$. No rotation is assumed for the BS and the LIS while $\alpha$-rad rotation is assumed for the MS. The objective of the system is to localize the MS and estimate its orientation by using the received signals at the MS with $N$ mmWave orthogonal frequency division multiplexing (OFDM) subcarriers. 

The propagation channel is composed of one direct path and one reflection path via the LIS. The direct channel between the BS and MS for the $n$-th subcarrier is expressed as 
\begin{align}\label{Channel_model_direct}
\bH_\text{B,M}[n] &=  \rho_\text{B,M} e^{-j2\pi\tau_\text{B,M} \frac{nB}{N}} \boldsymbol{\alpha}_{r}(\phi_\text{B,M}) \boldsymbol{\alpha}_{t}^H(\theta_\text{B,M}) , \nonumber\\ &\;\;\text{for} \; n = - (N-1)/2,\cdots, (N-1)/2,
\end{align}
where $\boldsymbol{\alpha}_{r}(\phi_\text{B,M}) \in \mathbb{C}^{N_\text{M} \times 1}$ and $\boldsymbol{\alpha}_{t}(\theta_\text{B,M}) \in \mathbb{C}^{N_\text{B} \times 1}$ are the antenna array response and steering vectors at the MS and BS, respectively. The $i$-th entry of $\boldsymbol{\alpha}_{r}(\phi_\text{B,M})$ and $\boldsymbol{\alpha}_{t}(\theta_\text{B,M})$ are $[\boldsymbol{\alpha}_{r}(\phi_\text{B,M})]_i = e^{j 2 \pi (i-1) \frac{d}{\lambda} \sin(\phi_\text{B,M})}$, $[\boldsymbol{\alpha}_{t}(\theta_\text{B,M})]_i = e^{j 2 \pi (i-1) \frac{d}{\lambda} \sin(\theta_\text{B,M})}$ with $d$ being the antenna element spacing\footnote{With notation reuse, $d$ also denotes element spacing in the LIS.}, $\lambda$ being the wavelength of the signal, and $\theta_\text{B,M}$ and $\phi_\text{B,M}$ being the AoD and AoA, respectively. $j = \sqrt{-1}$, $\tau_\text{B,M}$ is the ToA, $B$ is the overall bandwidth for all the subcarriers, and $B \ll f_c$\footnote{All the wavelengths $\lambda$'s of the subcarriers are nearly the same because of $B \ll f_c$.}, where $f_c$ is the center frequency. $\rho_\text{B,M} \in \mathbb{R}^+$ is the free-space path loss occurred in the direct link for all the subcarriers, and $(\cdot)^H$ denotes the conjugate transpose operation.

The two tandem channels ($\bH_\text{B,L}[n] \in \mathbb{C}^{N_\text{L} \times N_\text{B}}$ for the first hop and $\bH_\text{L,M}[n]\in \mathbb{C}^{N_\text{M} \times N_\text{L}}$ for the second hop) for the $n$-th subcarrier, which connect the BS to the MS via the LIS, are defined as 
\begin{equation}
 \bH_\text{B,L}[n]=  \rho_\text{B,L} e^{-j2\pi\tau_\text{B,L} \frac{nB}{N}}\boldsymbol{\alpha}_{r}(\phi_\text{B,L})\boldsymbol{\alpha}_{t}^H(\theta_\text{B,L}), 
\end{equation} 
and 
\begin{equation}
\bH_\text{L,M}[n] =  \rho_\text{L,M} e^{-j2\pi\tau_\text{L,M} \frac{nB}{N}} \boldsymbol{\alpha}_{r}(\phi_\text{L,M}) \boldsymbol{\alpha}_{t}^H(\theta_\text{L,M}), 
\end{equation} 
where the notations $\boldsymbol{\alpha}_{t}(\theta_\text{B,L})$, $\boldsymbol{\alpha}_{r}(\phi_\text{B,L})$, $\boldsymbol{\alpha}_{t}(\theta_\text{L,M})$, $\boldsymbol{\alpha}_{r}(\phi_\text{L,M})$, $ \rho_\text{B,L}$, $\rho_\text{L,M}$, $\tau_\text{B,L}$, and $\tau_\text{L,M}$ are defined in the same way as those in \eqref{Channel_model_direct}.


The entire channel, including both the line-of-sight (LoS) path and the non-line-of-sight (NLoS) path (i.e., the reflection path via the LIS), between the BS and the MS for the $n$-th subcarrier can be formulated as
\begin{equation}\label{LIS_Channel1}
\bH[n] = \bH_\text{B,M}[n] + \bH_\text{L,M}[n] \boldsymbol{\Omega} \bH_\text{B,L}[n],
\end{equation}  
where $\boldsymbol{\Omega} = \mathrm{diag}(\exp\{j \omega_1\} ,\cdots, \exp\{j \omega_{N_\text{L}}\} )  \in \mathbb{C}^{N_\text{L} \times N_\text{L}}$ is the phase control matrix at the LIS. It is a diagonal matrix with constant-modulus entries in the diagonal. 

Assuming that precoding $\bF$ is exploited at the BS and the positioning reference signal (PRS) $\bx[n]$ is transmitted over the $n$-th subcarrier, the downlink received signal is in the form of 
\begin{equation}
\by[n] = \sqrt{P}\bH[n] \bF \bx[n] +  \bn[n],
\end{equation} 
where each entry in the additive white noise $\bn[n]$ follows circularly-symmetric complex normal distribution $\mathcal{CN}(0, 2\sigma^2)$, and $P$ is the transmit power of the PRS. 

The geometric relationship among the BS, LIS, and MS is formulated as 
\begin{align}\label{Geometry_relationship}
&\tau_\text{B,M} = \|\bb -\bm\|_2/c, \nonumber\\
&\tau_\text{B,L} +\tau_\text{L,M} = \|\bb - \bl \|_2/c + \|\bm - \bl \|_2/c,\nonumber\\
&\theta_\text{B,M} = \arccos((m_x - b_x)/ \|\bb -\bm\|_2),\nonumber\\
&\theta_\text{B,L} = \arccos((l_x - b_x)/ \|\bl -\bb\|_2),\nonumber\\
&\theta_\text{L,M} = -\arccos(( m_x-l_x)/ \|\bl-\bm\|_2),\nonumber\\
&\phi_\text{B,M} = \pi+\arccos((m_x - b_x)/ \|\bb -\bm\|_2) - \alpha \nonumber\\
&\;\;\;\;\;\; =  \pi + \theta_\text{B,M} - \alpha, \nonumber\\
&\phi_\text{B,L} = -\pi+\arccos((l_x - b_x)/ \|\bb -\bl\|_2) =-\pi+ \theta_\text{B,L},\nonumber\\
&\phi_\text{L,M} = \pi - \arccos((m_x - l_x)/ \|\bl -\bm\|_2) - \alpha \nonumber\\
&\;\;\;\;= \pi +\theta_\text{L,M}-\alpha, \nonumber\\
&\rho_\text{B,M} =   (\|\bb -\bm\|_2)^{-\mu/2}, \nonumber\\
&\rho_\text{B,L} =  (\|\bb -\bl\|_2)^{-\mu/2}, \nonumber\\
&\rho_\text{L,M} =  (\|\bl -\bm\|_2)^{-\mu/2}, 
\end{align}
where $\bb = [b_x \; b_y]^T$, $\bl = [l_x \; l_y]^T$, and $\bm = [m_x \; m_y]^T$ are the centers of the BS, LIS, and MS, respectively, $\alpha$ is the orientation of the MS, $\mu$ is the path loss exponent, $c$ is the speed of light, and $\|\cdot\|_2$ stands for the Euclidean norm. Based on Fig.~\ref{System_Model}, we can further impose the following constraints on the channel angular parameters: 1) $\theta_\text{B,M},\theta_\text{B,L}\in (0, \pi/2)$, 2) $\theta_\text{L,M} \in (-\pi/2, 0)$, 3) $\phi_\text{B,L} \in (-\pi, -\pi/2)$, 4) $\phi_\text{B,M}, \phi_\text{L,M} \in (\pi/2, \pi)$, and 5) $\alpha \in (0, \pi/2)$. It should be noted that we consider far-field communications. Therefore, additional constraints are imposed to the number of LIS elements: $\frac{2 (N_\text{L} d)^2}{\lambda} < \| \bb -\bl\|$ and $\frac{2 (N_\text{L} d)^2}{\lambda} < \| \bl -\bm\|$, which can be summarized as $N_\text{L} < \frac{\sqrt{\lambda}}{\sqrt{2} d} \cdot \min\{ \sqrt{\| \bb -\bl\|}, \sqrt{ \| \bl -\bm\|}\}$. 

Under the condition that the positions of the BS and the LIS are known \textit{a priori}, the system can be virtually regarded as a two-LoS aided positioning system. Intuitively, better position estimation accuracy is expected compared to the scenario, which is a mixture of one LoS path and one NLoS path~\cite{Shahmansoori2018}. 

\section{Problem Formulation}
We use the two-stage approach to estimate the user coordinate and orientation. In the first stage, we estimate the channel parameters, defined as $\boldsymbol{\eta}= [\tau_\text{B,M}, \theta_\text{B,M}, \phi_\text{B,M}, \rho_\text{B,M}, \tau_\text{L,M}, \phi_\text{L,M}, \rho_\text{L,M}]^T$ with $(\cdot)^T$ denoting the transpose operation. The estimate of $\boldsymbol{\eta}$ can be presented in a general form as 
\begin{equation}
\hat{\boldsymbol{\eta}} = \boldsymbol{\eta} + \bw,
\end{equation}
where $\bw \sim \mathcal{CN}(\boldsymbol{0}, \boldsymbol{\Sigma})$ denotes the estimation error.   

Relying on the estimate $\hat{\boldsymbol{\eta}}$, we further obtain the MS's coordinate $\hat{\bm}$ and orientation $\hat{\alpha}$ via
\begin{align}
[\hat{\bm}, \hat{\alpha}] &= \argmax\limits_{[\bm, \alpha]} \; p(\hat{\boldsymbol{\eta}}|\boldsymbol{\eta}(\bm, \alpha)) \\
&= \argmin\limits_{[\bm, \alpha]}\; (\hat{\boldsymbol{\eta}} -\boldsymbol{\eta}(\bm, \alpha) )^T \boldsymbol{\Sigma}^{-1} (\hat{\boldsymbol{\eta}} -\boldsymbol{\eta}(\bm, \alpha)),
\end{align}
where $\boldsymbol{\eta}(\bm, \alpha)$ is a function of $\bm$ and $\alpha$, building the relationship among $\boldsymbol{\eta}$, $\bm$, and $\alpha$ detailed in~\eqref{Geometry_relationship}.

In practice, the estimate of $\boldsymbol{\eta}$ can be done via compressive sensing techniques, e.g., orthogonal matching pursuit (OMP)~\cite{Donoho2006}, basis pursuit (BP)~\cite{Donoho2006}, or approximate message passing (AMP)~\cite{Hannak2018} due to the inherent sparsity property of the mmWave MIMO channels~\cite{Rappaport2013}.

The goal of the LIS aided mmWave MIMO positioning system is to minimize the average distortion of the position estimation with Euclidean distance measure, i.e., 
\begin{equation}
\mathrm{var}(\hat{\bm}) = E [(m_x - \hat{m}_x)^2] + E [(m_y - \hat{m}_y)^2],
\end{equation}
and that of orientation estimation, i.e., 
\begin{equation}
\mathrm{var}(\hat{\alpha}) = E [(\alpha - \hat{\alpha})^2],
\end{equation}
where $E[\cdot]$ is the expectation operator. 

\section{Cram\'er Rao Lower Bounds}
In general, we aim at calculating the Cram\'er Rao lower bounds for the vector-valued unknown parameter $\boldsymbol{\zeta} = [m_x \; m_y \; \alpha]^T$. However, it is not straightforward to obtain them. We first calculate the Fisher information matrix (FIM) of $\boldsymbol{\eta}$ for the $n$-th subcarrier, defined as 
$\bar{\bJ}[n] \in \mathbb{R}^{7\times 7}$ with $[\bar{\bJ}[n] ]_{i,j} = \Psi_n(\eta_i, \eta_j) = \frac{P}{\sigma^2}\Re \{\frac{\partial \boldsymbol{\mu}^H[n]} {\partial \eta_i} \frac{ \partial \boldsymbol{\mu}[n]}{ \partial \eta_j} \}$, where $\boldsymbol{\mu}[n] = \sqrt{P}\bH[n] \bF \bx[n]$. The details on all the elements in $\bar{\bJ}_n$ are described in Appendix~\ref{FIM}.

\tcr{
\textbf{Observation 1}: According to~\eqref{J_tau_bm},~\eqref{J_theta_bm},~\eqref{J_phi_bm}, and~\eqref{J_rho_bm}, the estimate of channel parameters in the direct link is independent from the NLoS via the LIS. The estimation performance depends on the design of precoding matrix $\bF$ and PRS $\bx[n]$. }

\tcr{
\textbf{Observation 2}: According to~\eqref{J_tau_lm}, \eqref{J_phi_lm}, and \eqref{J_rho_lm}, the estimate of channel parameters in the indirect link is independent from the LoS. The estimation performance depends on $\beta[n]$, which is a function of $\bF$, PRS $\bx[n]$ and $\boldsymbol{\Omega}$, in the form of 
\begin{align}
\beta[n] &= \boldsymbol{\alpha}^H_{t}(\theta_\text{L,M}) \boldsymbol{\Omega}\boldsymbol{\alpha}_{r}(\phi_\text{B,L})\boldsymbol{\alpha}^H_{t}(\theta_\text{B,L})   \bF \bx[n]\nonumber\\
&= \left[\boldsymbol{\alpha}_{t}(\theta_\text{L,M}) \odot \boldsymbol{\alpha}^*_{r}(\phi_\text{B,L})\right]^H \boldsymbol{\omega} \boldsymbol{\alpha}^H_{t}(\theta_\text{B,L})   \bF \bx[n], \label{beta_in_FIM}
\end{align}
where $\boldsymbol{\Omega}=\mathrm{diag}(\boldsymbol{\omega})$ and $\odot$ denotes element-wise product. $|\beta[n]| \leq N_\text{L} | \boldsymbol{\alpha}^H_{t}(\theta_\text{B,L})   \bF \bx[n]|$. When $\omega_i = 2\pi(i-1) \frac{d}{\lambda} [\sin(\theta_\text{L,M}) - \sin(\phi_\text{B,L}) ]$, $|\beta[n]| = N_\text{L} | \boldsymbol{\alpha}^H_{t}(\theta_\text{B,L})   \bF \bx[n]|$. In other words, when $\bF$ and $\bx[n]$ are fixed, we can get the optimal estimate of channel parameters in the indirect link when the phase control matrix at LIS satisfies the following condition: $\omega_i = 2\pi(i-1) \frac{d}{\lambda} [\sin(\theta_\text{L,M}) - \sin(\phi_\text{B,L}) ]$.}

Then, we derive the Jacobian matrix $\bT_1$ with $[\bT_1]_{i,j} =  \partial \eta_i/ \partial \zeta_j$, detailed in the below:
\begin{align}
& \partial\tau_\text{B,M}/ \partial m_x =\frac{\cos(\theta_\text{B,M})}{c }, \nonumber\\
&\partial\theta_\text{B,M}/ \partial m_x  =  \partial\phi_\text{B,M}/ \partial m_x= -\frac{\sin(\theta_\text{B,M})}{\|\bb-\bm\|_2 },  \nonumber\\
&\partial \rho_\text{B,M}/ \partial m_x = - \mu/2 \|\bb-\bm\|_2 ^{-\mu/2-1}\cos(\theta_\text{B,M}) , \nonumber\\
&\partial\tau_\text{L,M}/ \partial m_x =\frac{\cos(\theta_{l,u} )}{c}, \partial\phi_\text{L,M}/ \partial u_x  = -\frac{\sin(\theta_\text{L,M})}{\|\bl-\bm\|_2 }, \nonumber
\end{align}
\begin{align}
&\partial \rho_\text{L,M}/ \partial m_x =- \mu/2 \|\bl-\bm\|_2 ^{-\mu/2-1}\cos(\theta_\text{L,M}) , \nonumber\\
&\partial\tau_\text{B,M}/ \partial m_y = \frac{\sin(\theta_\text{B,M})}{c }, \nonumber\\
& \partial\theta_\text{B,M}/ \partial m_y= \partial\phi_\text{B,M}/ \partial m_y= \frac{\cos(\theta_\text{B,M})}{\|\bb-\bm\|_2 },  \nonumber\\
&\partial \rho_\text{B,M}/ \partial m_y = - \mu/2 \|\bb-\bm\|_2 ^{-\mu/2-1} \sin(\theta_\text{B,M}), \nonumber\\
&\partial\tau_\text{L,M}/ \partial m_y = \frac{\sin(\theta_\text{L,M})}{c },  \partial\phi_\text{L,M}/ \partial m_y= \frac{\cos(\theta_\text{L,M})}{\|\bl-\bm\|_2 }, \nonumber\\
&\partial \rho_\text{L,M}/ \partial m_y = - \mu/2 \|\bl-\bm\|_2 ^{-\mu/2-1} \sin(\theta_\text{L,M}), \nonumber\\
&\partial\tau_\text{B,M}/ \partial \alpha = 0, \nonumber\\
&\partial\theta_\text{B,M}/ \partial \alpha = -1,  \partial\phi_\text{B,M}/ \partial \alpha = 1,\nonumber\\
&\partial\rho_\text{B,M}/ \partial \alpha = 0, \partial\tau_\text{L,M}/ \partial \alpha = 0, \nonumber\\
&\partial\phi_\text{L,M}/ \partial \alpha  = -1, \partial \rho_\text{L,M}/ \partial \alpha = 0. 
\end{align}
The FIM of $\boldsymbol{\zeta}$ for the $n$-th subcarrier is 
\begin{equation}
\tilde{\bJ}[n] = \bT_1 \bar{\bJ}[n] \bT_1^T,
\end{equation}
and by summing up all the contributions from the $N$ subcarriers, the FIM $\tilde{\bJ}$ of $\boldsymbol{\zeta}$ is in the form of 
\begin{equation}
\tilde{\bJ} = \sum_{n = -(N-1)/2}^{(N-1)/2} \tilde{\bJ}[n].
\end{equation}

The objective is to find the minimal theoretically achievable value for the standard deviation of positioning estimation error and orientation estimation error, which is the Cram\'er Rao lower bound, written as
\begin{equation}
\text{PEB} = \sqrt{\mathrm{tr}\{[\tilde{\bJ}^{-1}]_{1:2,1:2}\}} \leq \sqrt{\mathrm{var}(\hat{\bm})} ,
\end{equation}
and 
\begin{equation}
\text{OEB} = \sqrt{[\tilde{\bJ}^{-1}]_{3,3}} \leq \sqrt{\mathrm{var}(\hat{\alpha}) }.
\end{equation}

\section{Simulation Results}
The parameters are set up as follows: $(b_x, b_y )= (0,0)$, $(l_x, l_y) = (160/3, 80)$, $(m_x, m_y) =(80, 40)$, $\alpha = \pi/10$, $\mu = 2.08$ (path loss exponent), $N_\text{B} =128$, $N_\text{M} = 32$, $N = 31$, $B = 100$ MHz, $f_c = 60$ GHz, and $\alpha = \pi/10$. \tcr{According to the far field constraints, $N_\text{L} \leq 138$.} The signal-to-noise ratio (SNR) is defined as $\frac{P}{\sigma^2}$. For the purpose of comparison, we introduce a benchmark scenario~\cite{Shahmansoori2018} with one LoS and one NLoS path with the scatter located at $(160/3, 80)$. \tcr{For the calculation of theoretical performance limits, the location of the scatter is deterministic but unknown to the MS.} 
\subsection{Impact of the LIS: Phase Shifter}
\begin{figure}[t]
	\centering
	\includegraphics[width=0.85\linewidth]{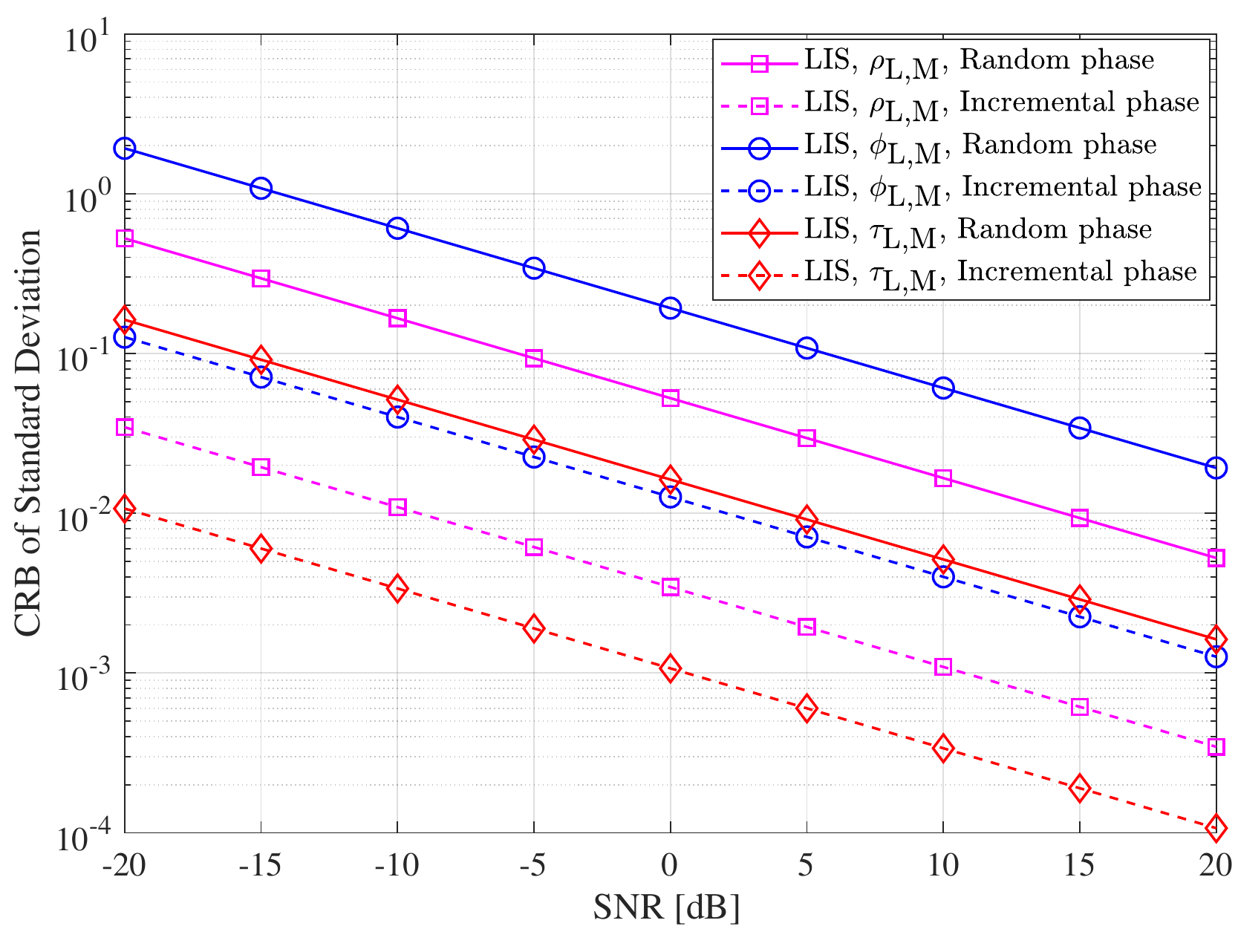}
	\caption{The impact of phases on CRB of  standard deviation of channel parameters in the reflection path with $N_\text{L} = 100$.}
	\label{Phase_effect_N_r_100}
\end{figure}

In this experiment, we set each element of $\bf = \bF\bx[n]$ as $e^{j \nu}$ with $\nu \sim \mathcal{U}(0\;\; 2\pi]$, $\forall n$. It shown in \eqref{beta_in_FIM} that if $\omega_i = 2\pi(i-1) \frac{d}{\lambda} [\sin(\theta_\text{L,M}) - \sin(\phi_\text{B,L}) ]$, the diagonal entries in FIM, e.g.,~\eqref{J_tau_lm}, \eqref{J_phi_lm}, and \eqref{J_rho_lm}, can achieve the maximum value, i.e., $\boldsymbol{\alpha}^H_{t}(\theta_\text{L,M}) \boldsymbol{\Omega}\boldsymbol{\alpha}_{r}(\phi_\text{B,L})$ in $\beta$ satisfies the following condition $\boldsymbol{\alpha}^H_{t}(\theta_\text{L,M}) \boldsymbol{\Omega}\boldsymbol{\alpha}_{r}(\phi_\text{B,L}) = N_\text{L}$ when $\omega_i = 2\pi(i-1) \frac{d}{\lambda} [\sin(\theta_\text{L,M}) - \sin(\phi_\text{B,L}) ]$ (labeled as ``Incremental phase'' in Fig.~\ref{Phase_effect_N_r_100}). This in turn provides a better estimation performance of NLoS channel parameters from the theoretical perspective, which is verified in Fig.~\ref{Phase_effect_N_r_100}. \tcr{``\text{LIS}'' in the legend stands for the studied LIS aided mmWave MIMO positioning system.} Note that all the simulation curves are obtained under the same randomly-generated $\bf$ and other parameters but only different phase conditions (benchmark scheme with random $\omega_i$ in $\boldsymbol{\Omega}$, labeled as ``Random phase'' in Fig.~\ref{Phase_effect_N_r_100}). \tcr{The incremental phase significantly outperforms the random phase (roughly $10$ times better in the current experiment)}. The optimal design of $\bf$ will bring better performance, and be left as our future work. 
\subsection{Impact of the LIS: Number of Elements}
\begin{figure}[t]
	\centering
	\includegraphics[width=0.85\linewidth]{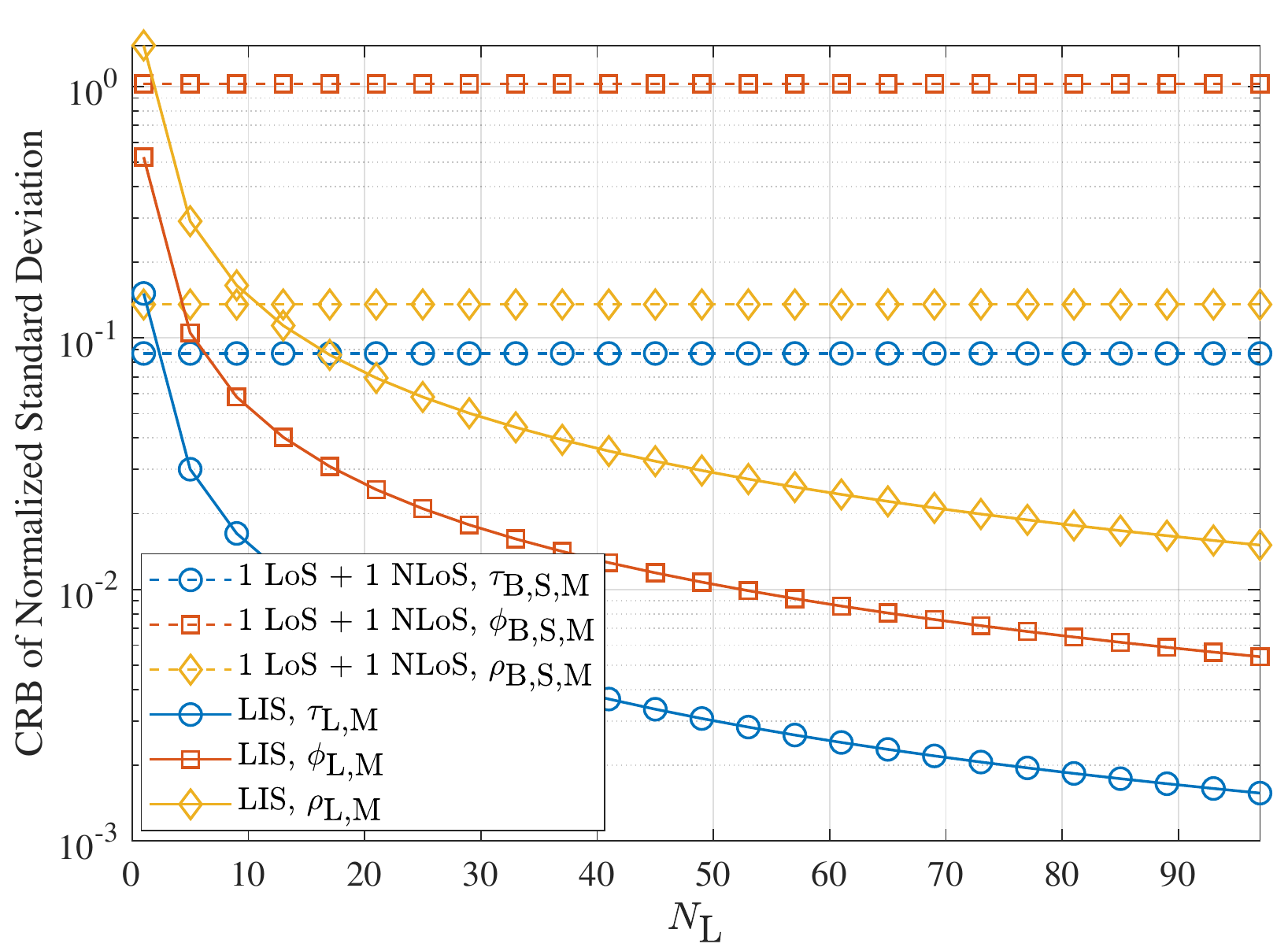}
	\caption{CRB of normalized standard deviation of channel parameters versus the number of LIS elements.}
	\label{Number_effect_SNR_5dB}
\end{figure}

In this experiment, we set $\omega_i = 2\pi(i-1) \frac{d}{\lambda} [\sin(\theta_\text{L,M}) - \sin(\phi_\text{B,L}) ]$ and SNR to 5 dB. The number of LIS elements plays a critical role on the estimation performance of channel parameters in the reflection path. The CRB of normalized standard deviation of $\tau_\text{L,M}$, $\phi_\text{L,M}$, and $\rho_\text{L,M}$ are inversely proportional to the number of elements in the LIS, e.g., $N_\text{L}$. As verified in Fig.~\ref{Number_effect_SNR_5dB}, the larger the number of elements in the LIS, the better the estimation performance of the channel parameters. \tcr{With around $100$ LIS elements, the estimation performance can be improved around $100$ times from the results shown in Fig.~\ref{Number_effect_SNR_5dB}.} In the legends, $\rho_\text{B,S,M}$, $\phi_\text{B,S,M}$, and $\tau_\text{B,S,M}$ denote the path loss, AoA, and ToA of the NLoS path via the scatter, respectively, for the benchmark scheme. Due to the fixed number of scatter, the estimation performance of the benchmark scheme stays unchanged. Note that the increase of elements does not have any great impact on the estimation accuracy of channel parameters of the direct path, which is not difficult to understand based on the FIM $\bar{\bJ}[n]$.

\subsection{PEB and OEB}
\begin{figure}[t]
	\centering
	\includegraphics[width=0.85\linewidth]{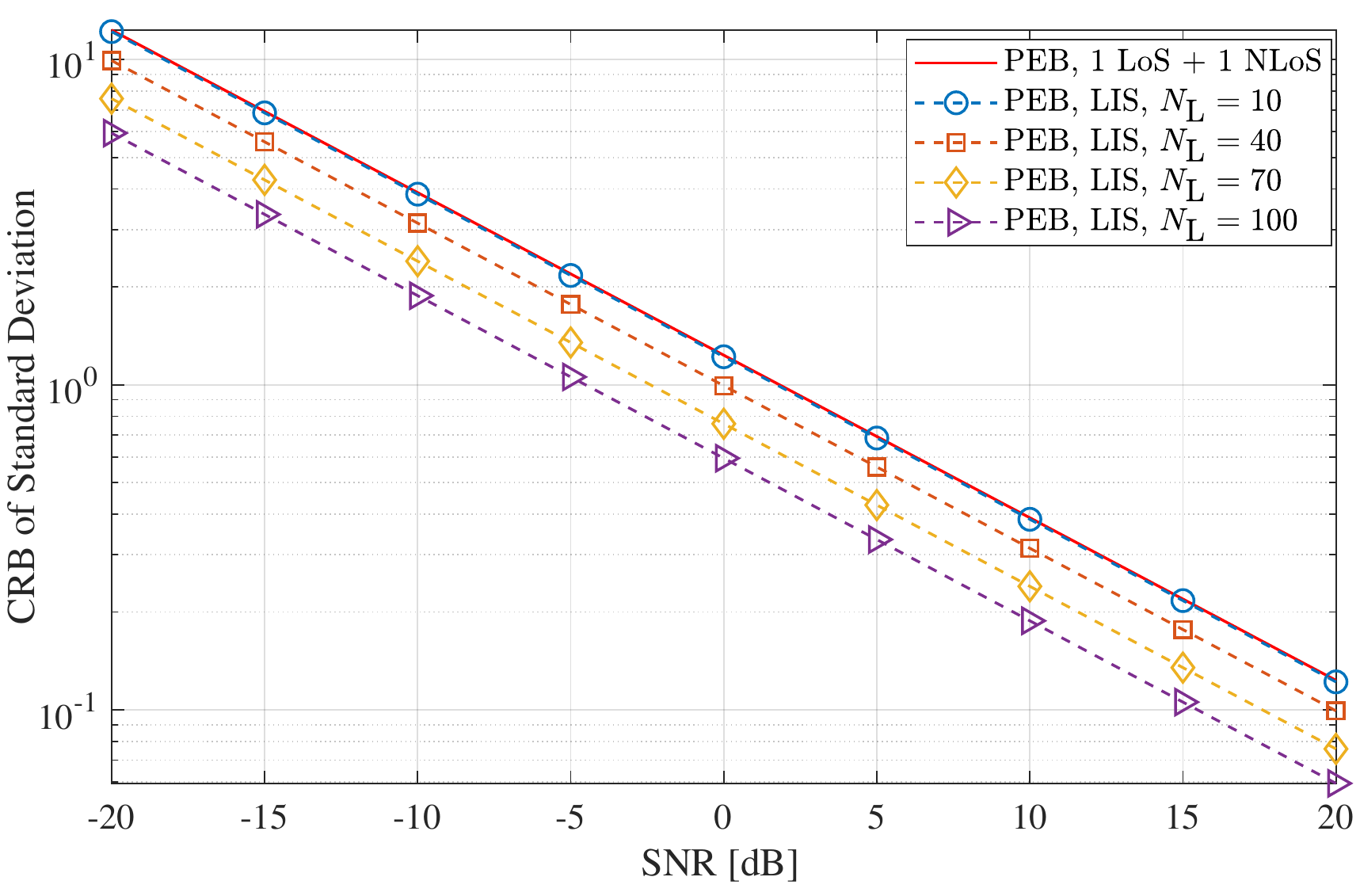}
	\caption{PEB versus the SNR.}
	\label{LIS_LOS_CRB_PEB_comparison}
\end{figure}

\begin{figure}[t]
	\centering
	\includegraphics[width=0.9\linewidth]{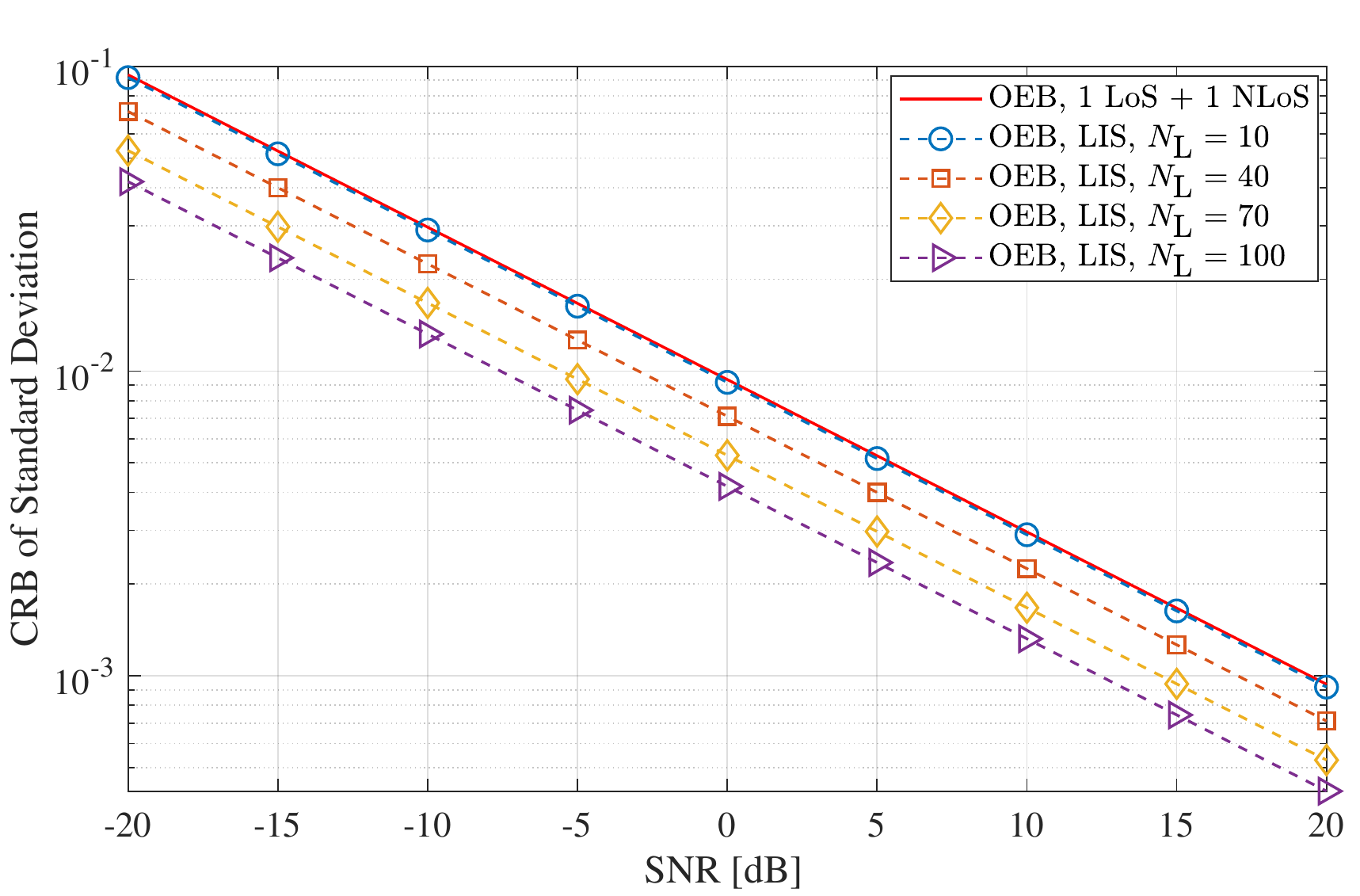}
	\caption{OEB versus the SNR.}
	\label{LIS_LOS_CRB_OEB_comparison}
\end{figure}

In this subsection, we evaluate the impact of the number of LIS elements on both the PEB and the OEB while fixing $\omega_i = 2\pi(i-1) \frac{d}{\lambda} [\sin(\theta_\text{L,M}) - \sin(\phi_\text{B,L}) ]$. As shown in Figs.~\ref{LIS_LOS_CRB_PEB_comparison} and~\ref{LIS_LOS_CRB_OEB_comparison}, the increase of elements improves the positioning performance. Even with the help of a $40$-element LIS, around 3 dB gain can be achieved when the OEB is at the level of $10^{-2}$. It should be noted that the performance enhancement mainly come from the improvement of the NLoS via the LIS. 

\section{Conclusions}
We have studied the fundamental limits of mmWave MIMO positioning with the aid of a LIS. The impact of the number of LIS elements and the phases of LIS elements on the estimation of channel parameters has been investigated. This in turn, helps the analyses on the positioning and orientation error bounds. The comparison has been made between the positioning system with and without the assistance of LIS to show the potential benefits brought by the introduction of LIS even with passive elements. Since the beamformer design at BS and phase shifter design at LIS play a critical role in the positioning, the joint consideration of them will be left as our future investigation.

\begin{appendices}
 \section{Derivation of FIM}\label{FIM}
The derivation of the FIM on all the channel parameters is shown as follows:
\tiny
\begin{align}
&\Psi_n(\tau_\text{B,M},\tau_\text{B,M}) =\frac{P N_\text{M} \rho^2_\text{B,M}}{ \sigma^2} \frac{(2 \pi n B)^2}{N^2} \| \boldsymbol{\alpha}^H_{t}(\theta_\text{B,M}) \bF\bx[n]\|_2^2 , \label{J_tau_bm}\\
&\Psi_n(\tau_\text{B,M}, \theta_\text{B,M}) =\frac{P \rho^2_\text{B,M}}{ \sigma^2}\Re\{ j 2 \pi \frac{nB}{N}  (\bx[n])^H \bF^H \boldsymbol{\alpha}_{t}(\theta_\text{B,M}) \dot{\boldsymbol{\alpha}}^H_{t}(\theta_\text{B,M}) \bF\bx[n] \}, \\
&\Psi_n(\tau_\text{B,M}, \phi_\text{B,M}) =\frac{P \rho^2_\text{B,M}}{ \sigma^2}  \nonumber \\ 
&\;\;\;\;\;\;\;\;\;\;\;\;\times \Re\{j 2 \pi \frac{nB}{N} (\bx[n])^H \bF^H  \boldsymbol{\alpha}_{t}(\theta_\text{B,M}) \boldsymbol{\alpha}^H_{r}(\phi_\text{B,M})\dot{\boldsymbol{\alpha}}_{r}(\phi_\text{B,M}) \boldsymbol{\alpha}^H_{t}(\theta_\text{B,M}) \bF \bx[n]\}, \\
&\Psi_n(\tau_\text{B,M}, \rho_\text{B,M}) =\frac{P \rho_\text{B,M}}{ \sigma^2} \| \boldsymbol{\alpha}^H_{t}(\theta_\text{B,M}) \bF\bx[n]\|_2^2 \Re\{j 2 \pi \frac{nB}{N} \} = 0, 
\end{align}
\begin{align}
&\Psi_n(\tau_\text{B,M}, \tau_\text{L,M}) =\frac{P \rho_\text{B,M} \rho_\text{B,L} \rho_\text{L,M}}{ \sigma^2 } \frac{(2 \pi n B)^2}{N^2} \nonumber \\
&\;\;\;\;\;\;\;\;\;\;\;\;\times \Re\{  \beta[n] \xi[n] (\bx[n])^H\bF^H  \boldsymbol{\alpha}_{t}(\theta_\text{B,M}) \boldsymbol{\alpha}^H_{r}(\phi_\text{B,M}) \boldsymbol{\alpha}_{r}(\phi_\text{L,M}) \}, \\
&\Psi_n(\tau_\text{B,M}, \phi_\text{L,M}) =\frac{P \rho_\text{B,M} \rho_\text{B,L} \rho_\text{L,M}}{ \sigma^2} \nonumber \\ 
&\;\;\;\;\;\;\;\;\;\;\;\;\times \Re\{ j 2 \pi \frac{nB}{N}\beta[n] \xi[n] (\bx[n])^H\bF^H  \boldsymbol{\alpha}_{t}(\theta_\text{B,M}) \boldsymbol{\alpha}^H_{r}(\phi_\text{B,M}) \dot{\boldsymbol{\alpha}}_{r}(\phi_\text{L,M})\}, \\
&\Psi_n(\tau_\text{B,M}, \rho_\text{L,M}) =\frac{P \rho_\text{B,M} \rho_\text{B,L} }{ \sigma^2} \nonumber \\ 
&\;\;\;\;\;\;\;\;\;\;\;\;\times  \Re\{ j 2 \pi \frac{nB}{N} \beta[n] \xi[n](\bx[n])^H\bF^H  \boldsymbol{\alpha}_{t}(\theta_\text{B,M}) \boldsymbol{\alpha}^H_{r}(\phi_\text{B,M}) \boldsymbol{\alpha}_{r}(\phi_\text{L,M})\}, \\
&\Psi_n(\theta_\text{B,M},\theta_\text{B,M}) = \frac{P N_\text{M} \rho^2_\text{B,M}}{ \sigma^2} \| \dot{\boldsymbol{\alpha}}^H_{t}(\theta_\text{B,M}) \bF\bx[n]\|_2^2,\label{J_theta_bm}\\
& \Psi_n(\theta_\text{B,M},\phi_\text{B,M})  = \frac{P \rho^2_\text{B,M}}{ \sigma^2} \nonumber \\ 
&\;\;\;\;\;\;\;\;\;\;\;\;\times  \Re\{ (\bx[n])^H \bF^H \dot{\boldsymbol{\alpha}}_{t}(\theta_\text{B,M}) \boldsymbol{\alpha}^H_{r}(\phi_\text{B,M})\dot{\boldsymbol{\alpha}}_{r}(\phi_\text{B,M}) \boldsymbol{\alpha}^H_{t}(\theta_\text{B,M}) \bF \bx[n] \}, \\
& \Psi_n(\theta_\text{B,M},\rho_\text{B,M}) = \frac{P N_\text{M}\rho_\text{B,M}}{ \sigma^2} \Re\{  (\bx[n])^H \bF^H \dot{\boldsymbol{\alpha}}_{t}(\theta_\text{B,M}) \boldsymbol{\alpha}^H_{t}(\theta_\text{B,M}) \bF \bx[n] \}, \\
&\Psi_n(\theta_\text{B,M},\tau_\text{L,M}) =\frac{P \rho_\text{B,M} \rho_\text{B,L} \rho_\text{L,M}}{ \sigma^2 } \nonumber \\  
&\;\;\;\;\;\;\;\;\;\;\;\;\times \Re\{ -j 2 \pi \frac{nB}{N}   \beta[n] \xi[n](\bx[n])^H\bF^H  \dot{\boldsymbol{\alpha}}_{t}(\theta_\text{B,M}) \boldsymbol{\alpha}^H_{r}(\phi_\text{B,M}) \boldsymbol{\alpha}_{r}(\phi_\text{L,M})\}, \\
& \Psi_n(\theta_\text{B,M},\phi_\text{L,M})  =\frac{P \rho_\text{B,M} \rho_\text{B,L} \rho_\text{L,M}}{ \sigma^2 }\nonumber \\  
&\;\;\;\;\;\;\;\;\;\;\;\;\times  \Re\{ \beta[n] \xi[n](\bx[n])^H\bF^H  \dot{\boldsymbol{\alpha}}_{t}(\theta_\text{B,M}) \boldsymbol{\alpha}^H_{r}(\phi_\text{B,M}) \dot{\boldsymbol{\alpha}}_{r}(\phi_\text{L,M})\}, \\
& \Psi_n(\theta_\text{B,M}, \rho_\text{L,M}) = \frac{P\rho_\text{B,M} \rho_\text{B,L}}{ \sigma^2  }\nonumber \\  
&\;\;\;\;\;\;\;\;\;\;\;\;\times   \Re\{ \beta[n] \xi[n] (\bx[n])^H\bF^H  \dot{\boldsymbol{\alpha}}_{t}(\theta_\text{B,M}) \boldsymbol{\alpha}^H_{r}(\phi_\text{B,M}) \boldsymbol{\alpha}_{r}(\phi_\text{L,M})\}, \\
& \Psi_n(\phi_\text{B,M}, \phi_\text{B,M}) = \frac{P  \rho^2_\text{B,M}}{ \sigma^2} \|  \dot{ \boldsymbol{\alpha}}_{r}(\phi_\text{B,M})  \boldsymbol{\alpha}^H_{t}(\theta_\text{B,M}) \bF\bx[n]\|_2^2,\label{J_phi_bm}\\
& \Psi_n(\phi_\text{B,M}, \rho_\text{B,M}) = \frac{P  \rho_\text{B,M}}{ \sigma^2}\nonumber \\  
&\;\;\;\;\;\;\;\;\;\;\;\;\times   \Re\{(\bx[n])^H \bF^H \boldsymbol{\alpha}_{t}(\theta_\text{B,M}) \dot{\boldsymbol{\alpha}}^H_{r}(\phi_\text{B,M})\boldsymbol{\alpha}_{r}(\phi_\text{B,M}) \boldsymbol{\alpha}^H_{t}(\theta_\text{B,M}) \bF \bx[n]\},\\
& \Psi_n(\phi_\text{B,M}, \tau_\text{L,M}) =\frac{P \rho_\text{B,M} \rho_\text{B,L} \rho_\text{L,M}}{ \sigma^2 }\nonumber \\  
&\;\;\;\;\;\;\;\;\;\;\;\;\times  \Re\{ -j 2 \pi \frac{nB}{N}\beta[n]  \xi[n](\bx[n])^H\bF^H  \boldsymbol{\alpha}_{t}(\theta_\text{B,M}) \dot{\boldsymbol{\alpha}}^H_{r}(\phi_\text{B,M}) \boldsymbol{\alpha}_{r}(\phi_\text{L,M})\}, \\
& \Psi_n(\phi_\text{B,M}, \phi_\text{L,M}) = \frac{P  \rho_\text{B,M} \rho_\text{B,L}\rho_\text{L,M}}{ \sigma^2 } \nonumber \\  
&\;\;\;\;\;\;\;\;\;\;\;\;\times \Re\{\beta[n] \xi[n] (\bx[n])^H\bF^H  \boldsymbol{\alpha}_{t}(\theta_\text{B,M}) \dot{ \boldsymbol{\alpha}}^H_{r}(\phi_\text{B,M}) \dot{\boldsymbol{\alpha}}_{r}(\phi_\text{L,M})\}, \\
& \Psi_n(\phi_\text{B,M}, \rho_\text{L,M}) =\frac{P \rho_\text{B,M} \rho_\text{B,L} }{ \sigma^2 } \nonumber \\  
&\;\;\;\;\;\;\;\;\;\;\;\;\times \Re\{  \beta[n] \xi[n] (\bx[n])^H\bF^H  \boldsymbol{\alpha}_{t}(\theta_\text{B,M}) \dot{\boldsymbol{\alpha}}^H_{r}(\phi_\text{B,M}) \boldsymbol{\alpha}_{r}(\phi_\text{L,M})\},\\
& \Psi_n(\rho_\text{B,M}, \rho_\text{B,M}) = \frac{P N_\text{M}}{ \sigma^2}  \| \boldsymbol{\alpha}^H_{t}(\theta_\text{B,M}) \bF\bx[n]\|_2^2,\label{J_rho_bm}\\
& \Psi_n(\rho_\text{B,M}, \tau_\text{L,M}) = \frac{P  \rho_\text{B,L} \rho_\text{L,M}}{ \sigma^2 } \nonumber \\  
&\;\;\;\;\;\;\;\;\;\;\;\;\times \Re\{ -j 2 \pi \frac{nB}{N}\beta[n] \xi[n](\bx[n])^H\bF^H  \boldsymbol{\alpha}_{t}(\theta_\text{B,M}) \boldsymbol{\alpha}^H_{r}(\phi_\text{B,M}) \boldsymbol{\alpha}_{r}(\phi_\text{L,M})\}, \\
& \Psi_n(\rho_\text{B,M}, \phi_\text{L,M}) =  \frac{P \rho_\text{B,L} \rho_\text{L,M}}{ \sigma^2}\nonumber \\  
&\;\;\;\;\;\;\;\;\;\;\;\;\times \Re\{ \beta[n] \xi[n](\bx[n])^H \bF^H \boldsymbol{\alpha}_{t}(\theta_\text{B,M}) \boldsymbol{\alpha}^H_{r}(\phi_\text{B,M})\dot{\boldsymbol{\alpha}}_{r}(\phi_\text{L,M})\}, \\
& \Psi_n(\rho_\text{B,M}, \rho_\text{L,M}) = \frac{P \rho_\text{B,L} }{ \sigma^2 }\nonumber \\
&\;\;\;\;\;\;\;\;\;\;\;\;\times  \Re\{   \beta[n] \xi[n](\bx[n])^H\bF^H  \boldsymbol{\alpha}_{t}(\theta_\text{B,M}) \boldsymbol{\alpha}^H_{r}(\phi_\text{B,M}) \boldsymbol{\alpha}_{r}(\phi_\text{L,M})\}, \\
&\Psi_n(\tau_\text{L,M},\tau_\text{L,M}) =\frac{P N_\text{M} \rho^2_\text{B,L} \rho^2_\text{L,M}}{ \sigma^2}  \frac{(2 \pi n B)^2 | \beta[n]|^2}{N^2} , \label{J_tau_lm}
\end{align}
\begin{align}
&\Psi_n(\tau_\text{L,M},\phi_\text{L,M}) =\frac{P  \rho^2_\text{B,L} \rho^2_\text{L,M}}{ \sigma^2 }\nonumber \\
&\;\;\;\;\;\;\;\;\;\;\;\;\times  \Re\{ j 2 \pi \frac{nB}{N} (\beta[n])^* \boldsymbol{\alpha}^H_{r}(\phi_\text{L,M})\dot{\boldsymbol{\alpha}}_{r}(\phi_\text{L,M})\boldsymbol{\Omega}\boldsymbol{\alpha}_{r}(\phi_\text{B,L})\boldsymbol{\alpha}^H_{t}(\theta_\text{B,L})   \bF \bx[n]\}, \\
&\Psi_n(\tau_\text{L,M},\rho_\text{L,M})  =  \frac{P \rho^2_\text{B,L} \rho_\text{L,M} | \beta[n]|^2}{ \sigma^2 }  \Re\{j 2 \pi \frac{nB}{N} \} = 0,  \\
&\Psi_n(\phi_\text{L,M},\phi_\text{L,M}) =  \frac{P  \rho^2_\text{B,L} \rho^2_\text{L,M}}{ \sigma^2 }   \| \beta[n] \dot{\boldsymbol{\alpha}}_{r}(\phi_\text{L,M})  \|_2^2,  \label{J_phi_lm}\\
&\Psi_n(\phi_\text{L,M},\rho_\text{L,M}) =  \frac{P \rho^2_\text{B,L} \rho_\text{L,M} }{ \sigma^2 } \nonumber \\  
&\;\;\;\;\;\;\;\;\;\;\;\;\times   \Re\{ (\beta[n])^* \boldsymbol{\alpha}^H_{r}(\phi_\text{L,M}) \dot{\boldsymbol{\alpha}}_{r}(\phi_\text{L,M})\boldsymbol{\Omega}\boldsymbol{\alpha}_{r}(\phi_\text{B,L})\boldsymbol{\alpha}^H_{t}(\theta_\text{B,L})   \bF \bx[n]\}, \\
&\Psi_n(\rho_\text{L,M},\rho_\text{L,M}) =\frac{PN_\text{M} \rho^2_\text{B,L} | \beta[n]|^2}{ \sigma^2 }  , \label{J_rho_lm}
\end{align} 
\normalsize
where $\dot{\boldsymbol{\alpha}}_{t}(\theta_\text{B,M})  =  \partial \boldsymbol{\alpha}_{t}(\theta_\text{B,M}) / \partial \theta_\text{B,M} =\bD_t(\theta_\text{B,M}) \alpha_{t}(\theta_\text{B,M}) $, $\dot{\boldsymbol{\alpha}}_{r}(\phi_\text{B,M}) =  \partial \boldsymbol{\alpha}_{r}(\phi_\text{B,M}) / \partial \phi_\text{B,M} = \bD_r(\phi_\text{B,M})\alpha_{r}(\phi_\text{B,M})$, $ \dot{\boldsymbol{\alpha}}_{r}(\phi_\text{L,M}) =  \partial \boldsymbol{\alpha}_{r}(\phi_\text{L,M}) / \partial \phi_\text{L,M}  = \bD_r(\phi_\text{L,M})\alpha_{r}(\phi_\text{L,M}) $ with $\bD_t(\theta_\text{B,M}) = j 2 \pi \frac{d}{\lambda} \cos(\theta_\text{B,M}) \mathrm{diag}(0, 1, \cdots,  i , \cdots, N_\text{B}-1)$, $\bD_r(\phi_\text{B,M}) = j 2 \pi \frac{d}{\lambda} \cos(\phi_\text{B,M}) \mathrm{diag}(0, 1, \cdots,  i , \cdots,N_\text{M}-1)$, $\bD_r(\phi_\text{L,M}) = j 2 \pi \frac{d}{\lambda} \cos(\phi_\text{L,M}) \mathrm{diag}(0, 1, \cdots,  i, \cdots, N_\text{M}-1)$, $\xi[n] = e^{-j[2\pi (\tau_\text{B,L}+\tau_\text{L,M}- \tau_\text{B,M}) \frac{nB}{N}]}$, and $\beta[n] = \boldsymbol{\alpha}^H_{t}(\theta_\text{L,M}) \boldsymbol{\Omega}\boldsymbol{\alpha}_{r}(\phi_\text{B,L})\boldsymbol{\alpha}^H_{t}(\theta_\text{B,L})   \bF \bx[n]$.

\end{appendices}

\bibliographystyle{IEEEtran}
\bibliography{IEEEabrv,Ref}

\end{document}